# Vector OFDM Transmission over Non-Gaussian Power Line Communication Channels

Bamidele Adebisi, *Senior Member, IEEE,* Khaled M. Rabie, *Member, IEEE,* Augustine Ikpehai, *Student Member, IEEE,* Cinna Soltanpur, *Member, IEEE,* and Andrew Wells, *Member, IEEE*

*Abstract*—Most of the recent power line communication (PLC) systems and standards, both narrow-band and broadband, are based on orthogonal frequency-division multiplexing (OFDM). This multiplexing scheme however suffers from the high peak-to-average power ratio (PAPR) which can considerably impact the energy efficiency, size and cost of PLC modems as well as cause electromagnetic compatibility (EMC) issues. This paper investigates the performance of vector OFDM (VOFDM), which has inherently better PAPR properties, over non-Gaussian broadband PLC channels equipped with two nonlinear preprocessors at the receiver. In addition, the low PAPR property of the VOFDM system is exploited to further enhance the efficiency of the nonlinear preprocessors. The achievable gains are studied in terms of the complementary cumulative distribution function of the PAPR, probability of noise detection error and the signal-to-noise ratio at the output of the nonlinear preprocessors. For comparison's sake, the performance of conventional OFDM systems is also presented throughout the paper. Results reveal that the proposed system is able to provide up to 2 dB saving in the transmit power relative to the conventional OFDM under same system conditions, which eventually also translates into a system that is more resilient to EMC limits, reduced cost and size of PLC modems. It is also shown that the achievable gains become more significant as the vector block (VB) size of the VOFDM system is increased.

## I. INTRODUCTION

IN addition to their traditional distribution of electricity, power lines have evolved into a communication medium for many applications in the areas of home-networking and smart grids [1]–[3]; this is also known as power line communication (PLC). This technology however is faced with several challenges such as high cable attenuation, frequency-selectivity, non-Gaussian noise as well as the limited transmit power restrictions to comply with electromagnetic compatibility (EMC) regulations [4]–[6]. To cope with these, several techniques have been reported in the literature including cooperative relaying, multiple-input multiple-output (MIMO) schemes, and very recently, energy harvesting in PLC systems [7]–[10].

Orthogonal frequency-division multiplexing (OFDM) has been widely used as the main transmission technique in most narrow-band and broadband PLC systems for its ability to combat frequency-selectivity as well as the non-Gaussian interference [11]–[13]. In addition, other merits of OFDM include its high spectral efficiency, adaptive power allocation and low implementation complexities through the use of inverse fast Fourier transform (IFFT) and fast Fourier transform (FFT). Despite the above qualities and its wide acceptance, one major drawback of conventional OFDM remains its high peak-to-average power ratio (PAPR) arising from the parallel processing of symbols by the IFFT, which can seriously affect the spectral efficiency and energy efficiency of OFDM-based PLC systems [13]–[15]. To reduce PAPR, different techniques have been introduced such as partial transmit sequence (PTS) and selective mapping (SLM) [13], [16]. These techniques however require side information transmission which can be energy inefficient and challenging to implement in practice, especially over non-Gaussian PLC channels.

In order to tackle this issue, we propose vector OFDM (VOFDM), which has inherently good PAPR features [17]–[19], for the non-Gaussian broadband PLC channel[1]. VOFDM can also offer more flexibility in system design, serving as a bridge connecting conventional OFDM and single-carrier frequency domain equalization. Two nonlinear preprocessors are implemented at the receiver to further improve the performance of the VOFDM system, namely nulling and clipping. It is worthwhile noting that numerous studies have investigated the performance of nulling and clipping in conventional OFDM systems, see e.g. [21]–[23] and the references therein. We evaluate the system performance in terms of the complementary cumulative distribution function (CCDF) of the PAPR, the probability of noise detection error and the output SNR. Results show that the proposed scheme can provide up to 2 dB transmit power savings compared to conventional OFDM to meet the same performance requirement. This reduction in the transmit power can consequently minimize the electromagnetic emissions from power lines, and also reduce cost and size of PLC modems as will be discussed later in more details. Furthermore, results demonstrate that the achievable gains will increase as we increase the number of vector blocks (VBs) of VOFDM and that the VOFDM-nulling system can generally offer more significant improvements than VOFDM with clipping.

The rest of the paper is structured as follows. Section II presents and discusses previous research, and Section III describes the system model. In Section IV, we examine the

B Adebisi, K. M. Rabie and A. Ikpehai are with the school of Electrical Engineering, Manchester Metropolitan University, Manchester, M15 6BH, UK. (e-mails: b.adebisi@mmu.ac.uk; k.rabie@mmu.ac.uk; augustine.ikpehai@stu.mmu.ac.uk).
C. Soltanpur is with the School of Electrical and Computer Engineering, University of Oklahoma, Norman, OK 73019 USA. (e-mail: cinna@ou.edu).
A. Wells is with Jaguar Land Rover Limited, Warwick CV35 0RR, U.K. (email: awells@jaguarlandrover.com).

[1]Note that VOFDM with Masreliez filtering has recently been investigated over non-Gaussian PLC channels in [20].

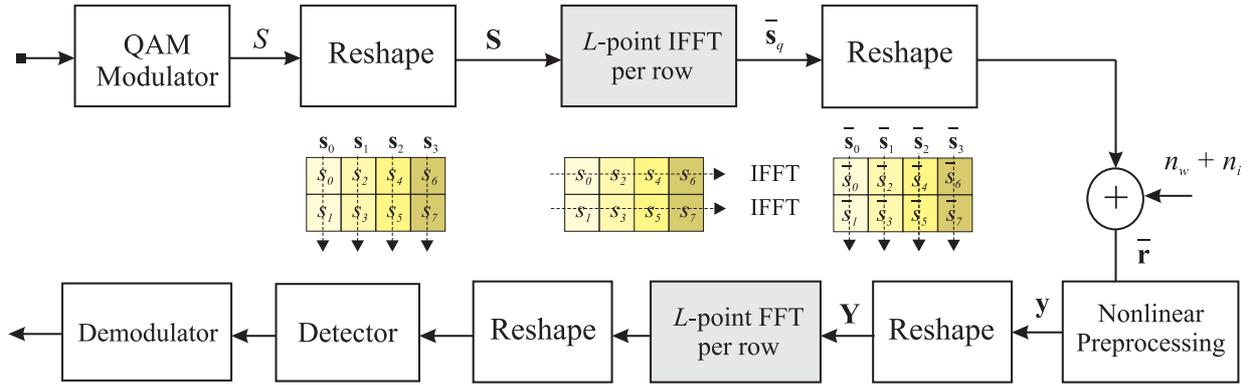

Figure 1: System diagram of the VOFDM system with nonlinear preprocessing at the receiver over non-Gaussian channels.

CCDF of the PAPR and discuss the probability of noise detection error for both VOFDM and conventional OFDM systems. The output SNR of the VOFDM system with nulling and clipping nonlinear preprocessors is studied in Section V. Section VI addresses the threshold optimization problem of the nonlinear preprocessors and presents analytical and simulated results. Section VII provides a brief comparison between VOFDM and conventional OFDM and highlights some practical implementation issues of the proposed system. Finally, Section VIII concludes the paper.

## II. Related Work and Our Contributions

Similar to many PLC solutions originated in the wireless domain, VOFDM was first introduced in the context of wireless communications by Xia in [24] to reduce the size of FFT, IFFT and the cycle prefix overhead. This was followed by many studies investigating different aspects of VOFDM systems. For instance, in 2005, Zhang et al. [25] analyzed some practical issues of VOFDM such as guard-band settings, synchronization and time estimation. In 2010, Han et al. [26] showed that the performances of different VBs in VOFDM can differ considerably when the maximum likelihood receiver is deployed. In addition, to overcome this and ensure consistent performance over all VBs, the authors proposed a new constellation rotation technique. Later in 2011, Cheng et al. [18] studied the performance of VOFDM in terms of diversity and coding gains over multipath Rayleigh fading channels with a maximum likelihood receiver. In 2012, Li et al. [15] proposed linear receivers for VOFDM such as zero-forcing (ZF) and minimum-mean square error (MMSE) receivers. Two years later, the authors of [27] investigated thoroughly the performance of VOFDM over fast fading channels. Very recently, Ngebani et al. [17] explored phase noise in VOFDM systems and proposed two algorithms to estimate and mitigate this noise using linear MMSE receivers.

All the above work, however, has been limited to additive white Gaussian noise (AWGN) wireless systems. In contrast, and to the best of our knowledge, this paper studies for the first time VOFDM over the non-Gaussian broadband PLC channel and then establishes a relationship between the low PAPR property of VOFDM and improving the noise cancellation at the PLC receiver. Note that in our evaluations, PLC noise is characterized using the Bernoulli-Gaussian noise model, [28], as it is the most widely used in evaluating the performance of broadband PLC systems, see e.g. [29]–[31].

The contributions of this work are as follows. First, we investigate the PAPR performance of VOFDM and relate its influence on the noise cancellation process at the receiver. Second, we examine the probability of noise detection error, the output SNR and the optimized system performances of the VOFDM-nulling and VOFDM-clipping systems. In addition, some practical implementation issues of the proposed system are also briefly discussed in comparison to conventional OFDM in terms of complexity, cost, energy efficiency and electromagnetic compatibility.

## III. System Model

In Fig. 1, we illustrate the block diagram of the system under consideration. This figure shows the transmitter and receiver sides of the VOFDM system. First, the information bits are mapped using quadrature-amplitude modulation (QAM) to produce base-band QAM symbols denoted as $S$. A sequence of $N$ modulated symbols is then column-wise blocked to $L$ vectors each of length $M$, i.e. $N = ML$. These vectors will be referred to as vector blocks (VBs) and the $l^{th}$ VB can be represented as

$$\mathbf{S}_l = [S_{lM}, S_{lM+1}, \ldots S_{lM+M-1}]^T \quad l = 0, 1, \ldots, L-1 \tag{1}$$

This can be written in a matrix format of $M$ rows and $L$ columns as

$$\mathbf{S} = \begin{pmatrix} S_0 & S_M & S_{2M} & \cdots & S_{(L-1)M} \\ S_1 & S_{M+1} & S_{2M+1} & \cdots & S_{(L-1)M+1} \\ \vdots & \vdots & \vdots & \ddots & \vdots \\ S_{M-1} & S_{2M-1} & S_{3M-1} & \cdots & S_{LM-1} \end{pmatrix} \tag{2}$$

After that, an IFFT of size $L$ is performed over the $M$ VBs component-wise as illustrated in the example in Fig. 1 (for $M = 2$ and $L = 4$). The VOFDM time-domain signal after the IFFT can then be expressed as

$$\bar{\mathbf{s}}_q = \frac{1}{\sqrt{L}} \sum_{l=0}^{L-1} \mathbf{S}_l \exp\left(\frac{j2\pi ql}{L}\right), \quad q = 0, 1, \ldots, L-1 \quad (3)$$

which can also be written in a vector form as

$$\bar{\mathbf{s}}_q = [\bar{s}_{qM}, \, \bar{s}_{qM+1}, \, \ldots, \bar{s}_{qM+M-1}]^T \quad q = 0, 1, \ldots, L-1 \quad (4)$$

The next step is to reshape the vector in (4) to yield a vector of length $N$, that is

$$\bar{\mathbf{s}} = \left[\bar{\mathbf{s}}_0^T, \bar{\mathbf{s}}_1^T, \ldots, \bar{\mathbf{s}}_{L-1}^T\right] = [\bar{s}_0, \bar{s}_1, \ldots, \bar{s}_{N-1}] \quad (5)$$

The corresponding PAPR of this signal is calculated as

$$\text{PAPR} = \frac{\max\left(|\bar{s}_k|^2\right)}{\mathbb{E}\left[|\bar{s}_k|^2\right]}, \quad k = 0, 1, \ldots, N-1 \quad (6)$$

where $\max(\cdot)$ denotes the maximum argument, $|\cdot|$ is the absolute value operator and $\mathbb{E}[\cdot]$ is the expectation operator.

The VOFDM signal is then transmitted over the PLC channel where it becomes contaminated with the PLC noise (consisting of background and impulsive components). Therefore, the received signal can be written in the following form

$$\bar{r}_k = \bar{s}_k + n_{w,k} + n_{i,k}, \quad k = 0, 1, \ldots, N-1 \quad (7)$$

or in a vector form as

$$\bar{\mathbf{r}} = [\bar{r}_0, \bar{r}_1, \ldots, \bar{r}_{N-1}]^T \quad (8)$$

where $\bar{r}_k$ is the received signal, and $n_w$ and $n_i$ represent the background and impulsive noise components, respectively. The Bernoulli-Gaussian noise model is used here to characterize both the background and impulsive noise, in which impulsive noise is generated as [28], [32]

$$n_{i,k} = \mathbf{b}\, g_k, \quad k = 0, 1, 2, \ldots, N-1 \quad (9)$$

where $g_k$ is complex white Gaussian noise with mean zero and b is the Bernoulli process with probability $\Pr(\mathbf{b} = 1) = p$, and $p$ is the probability occurrence of impulsive noise. Therefore, the probability density function (PDF) of the total noise $n_t = n_w + n_i$, can be expressed as

$$P_{n_t}(n_t) = \sum_{m=0}^{1} p_m \, \mathcal{G}\left(n_t, 0, \sigma_m^2\right) \quad (10)$$

where $\mathcal{G}(\cdot)$ is the Gaussian PDF given as $\mathcal{G}\left(x, \mu, \sigma_x^2\right) = \frac{1}{\sqrt{2\pi\sigma_x^2}} \exp\left(-\frac{(x-\mu)^2}{2\sigma_x^2}\right)$, $p_0 = (1-p)$, $p_1 = p$, $\sigma_0^2 = \sigma_w^2$ and $\sigma_1^2 = \sigma_w^2 + \sigma_i^2$. The variances $\sigma_w^2$ and $\sigma_i^2$ denote the background and impulsive noise powers and define the input SNR and signal-to-impulsive noise ratio (SINR), respectively, as $\text{SNR} = 10 \log_{10}\left(\frac{\sigma_s^2}{\sigma_w^2}\right)$ and $\text{SINR} = 10 \log_{10}\left(\frac{\sigma_s^2}{\sigma_i^2}\right)$, and $\sigma_s^2$ is the transmitted signal variance.

Commonly, to reduce the effect of noise in PLC systems, nonlinear preprocessors are implemented at the front-end of the receiver. Therefore, the received signal, $\bar{r}_k$, is now passed through a nonlinear preprocessor (either nulling or clipping) where the incoming signal is processed when it exceeds a predetermined threshold value.

- Nulling: in this scheme the received signal is set to zero when it exceeds the nulling threshold $(T_n)$. The principle of this device is

$$y_k = \begin{cases} \bar{r}_k, & |\bar{r}_k| \leq T_n \\ 0, & |\bar{r}_k| > T_n \end{cases} \quad k = 0, 1, \ldots, N-1 \quad (11)$$

where $y_k$ is the output of the nulling device.

- Clipping: the received signal in this configuration is clipped when it exceeds the clipping threshold $(T_c)$. The principle of this device is given mathematically as

$$y_k = \begin{cases} \bar{r}_k, & |\bar{r}_k| \leq T_c \\ T_c \exp\left(j \arg(\bar{r}_k)\right), & |\bar{r}_k| > T_c \end{cases} \quad (12)$$

where $y_k$ is the output of the clipper and $\arg(x)$ returns the argument of $x$.

After that, we column-wise block $\mathbf{y} = \{y_0, y_1, \ldots, y_{N-1}\}$ to an $M \times L$ matrix as follows

$$\mathbf{Y} = \begin{pmatrix} y_0 & y_M & y_{2M} & \cdots & y_{(L-1)M} \\ y_1 & y_{M+1} & y_{2M+1} & \cdots & y_{(L-1)M+1} \\ \vdots & \vdots & \vdots & \ddots & \vdots \\ y_{M-1} & y_{2M-1} & y_{3M-1} & \cdots & y_{LM-1} \end{pmatrix} \quad (13)$$

and then take the FFT over every row to produce the frequency-domain signal. This matrix is then reshaped to produce a $1 \times N$-size vector before performing the baseband demodulation and decision.

As mentioned earlier, one of the most attractive features of VOFDM over conventional OFDM is its low PAPR performance. This feature is exploited in this work to make PLC systems more robust. Therefore, a brief review of the PAPR properties of VOFDM is crucial to have an insight into its performance and to establish a relationship between PAPR reduction and the probability of noise detection error.

## IV. CCDF of VOFDM and Probability of Noise Detection Error

The CCDF is defined basically as the probability that the PAPR of the VOFDM symbol exceeds a certain threshold value, $\text{PAPR}_o$. That is

$$\text{CCDF} = 1 - \Pr\{\text{PAPR} \leq \text{PAPR}_o\}. \quad (14)$$

To illustrate the impact of the VBs on the PAPR performance of the VOFDM approach, we plot in Fig. 2 the PAPR performance versus the number of VBs for several values of the CCDF when $N = 256$ sub-carriers. It is seen from this figure that, for all the CCDF values considered, VOFDM always has better PAPR performance than conventional OFDM (i.e. $M = 1$), and that increasing the number of VBs will always further reduce PAPR. This is because VOFDM uses smaller IFFT size to generate its signal in comparison to conventional OFDM. The other observation one can see is that the PAPR reduction is more significant in low CCDF values and this becomes less pronounced as CCDF is increased. For instance, the PAPR reduction gain obtained when CCDF = 0.001 at $M = 64$ is around 5 dB relative to the conventional OFDM system whereas only about 2 dB gain is attained when CCDF = 0.5 at the same value of $M$. The other remark on these results is that when $M$ is very large, e.g. $M = 64$, the PAPR performance is equal for all the considered CCDF values. This PAPR reduction in conjunction with nonlinear preprocessing at the receiver will allow more accurate detection of the noise. It should be noted that even though increasing $M$ will increase the computational complexity, this is not very challenging to implement considering the advanced super-fast and low-power chips available today.

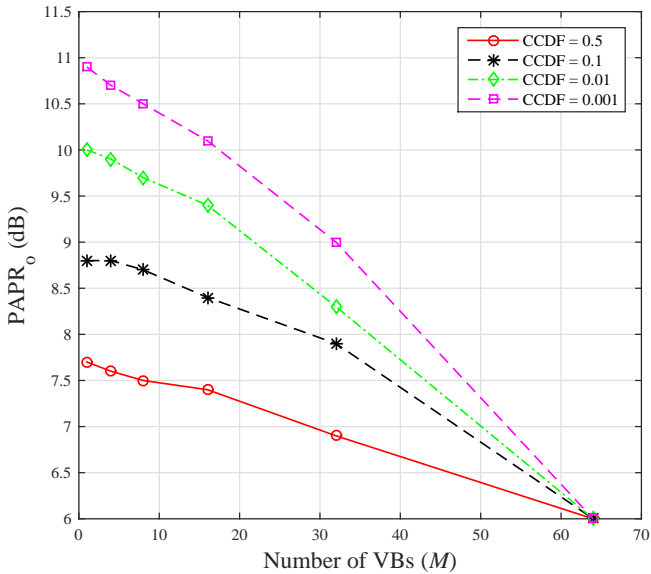

Figure 2: PAPR performance of the VOFDM system as a function of the number of VBs for several CCDF values. ($M = 1$ represents the conventional OFDM system).

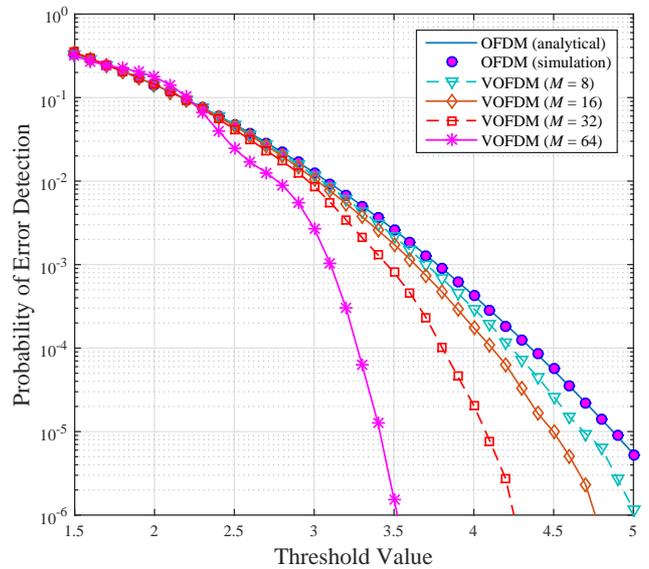

Figure 3: Probability of detection error performance versus the threshold value for the VOFDM system with various values of VBs when the input SNR = 25 dB.

We now look into the influence of reducing PAPR on the probability of noise detection error, $P_{de}$. This probability is defined as the probability that the amplitude of the VOFDM signal exceeds a predetermined threshold value when it is unaffected by noise, and is calculated as

$$P_{de} = \Pr\left(|\bar{r}_k| > T_i | \mathcal{H}_0\right) \Pr\left(\mathcal{H}_0\right) \tag{15}$$

where $i \in \{n, c\}$ and the null hypothesis $\mathcal{H}_0$ denotes the absence of impulsive noise.

Fig. 3 depicts some results for the probability of detection error for the VOFDM system with several VB sizes. It is evident that the VOFDM-based system always has better performance compared to conventional OFDM and this gain increases as we increase the number of VBs. It is also noticeable that when the threshold is low, both OFDM and VOFDM systems behave similarly regardless of the number of VBs deployed. However, as the threshold is increased, the gap between the two systems becomes larger. For more communication performance metric, we next evaluate to the output SNR performance and highlight the transmit power savings attainable with the proposed approach.

## V. Output SNR Performance

In this section, we investigate the transmit power savings obtained with the proposed system. We therefore consider the SNR at the output of the nonlinear preprocessors which can be found as [23]

$$\gamma_{o1} = \frac{\mathbb{E}\left[|R_1\,\bar{s}_k|^2\right]}{\mathbb{E}\left[|y_k - R_1\,\bar{s}_k|^2\right]} \tag{16}$$

where $R_1$ is a real constant chosen as $R_1 = (1/2)\,\mathbb{E}\left[|y_k\bar{s}_k^*|^2\right]$.

The system parameters used in this section onward are: $N = 256$ sub-carriers, input SNR = 25 dB, SINR = $-15$ dB and $p = 0.01$. To visualize the important impact of the threshold value on the system performance, we plot in Fig. 4 the output SNR of the proposed system with respect to the threshold value for both nulling and clipping cases with various VB sizes. For comparison, the output SNR of conventional OFDM is also included on this plot, the analytical results of which, for both nulling and clipping, can be calculated using

$$\gamma_{o2} = \frac{2R_2^2}{E_o - 2R_2^2} \tag{17}$$

where

$$R_2 = 1 - \sum_{i \in \{0,1\}} p_i\left[\exp\left(-\frac{T^2}{2\left(1 + \sigma_i^2\right)}\right) + T\,\Xi\right], \tag{18}$$

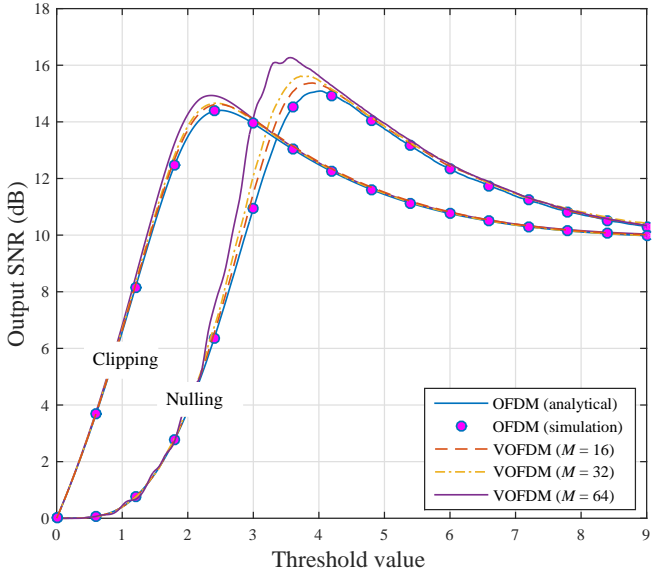

Figure 4: Output SNR performance of the proposed system versus the threshold value with nulling and clipping nonlinear preprocessing.

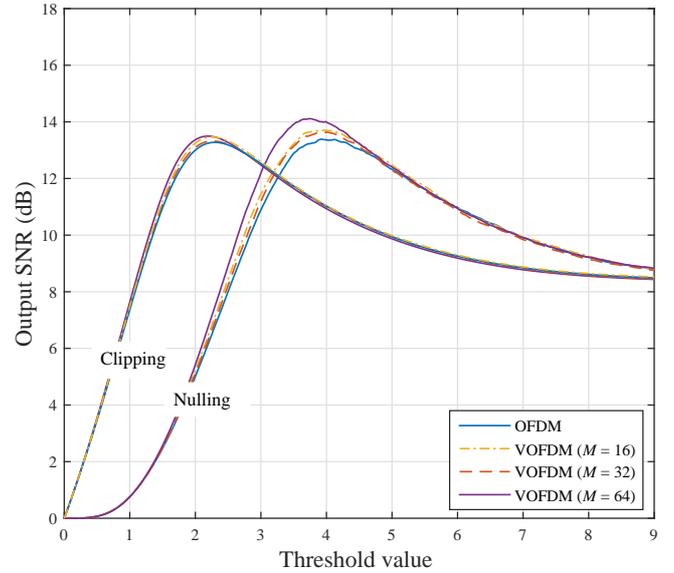

Figure 5: Output SNR performance of the proposed system versus the threshold value with nulling and clipping nonlinear preprocessing over a frequency selective channel.

$$E_o = 2 + 2 \sum_{i \in \{0,1\}} p_i \left( \sigma_i^2 - \Gamma \right) \exp \left( - \frac{T^2}{2 \left( 1 + \sigma_i^2 \right)} \right) \quad (19)$$

and $E_o$ is the total signal power at the output of the nonlinear preprocessor. For the nulling-based system $\Xi = \frac{T}{2\left(1+\sigma_i^2\right)} \exp\left(-\frac{T^2}{2\left(1+\sigma_i^2\right)}\right)$ and $\Gamma = 1 + \sigma_i^2$ whereas for the clipping-based scheme $\Xi = -\sqrt{\frac{\pi}{2\left(1+\sigma_i^2\right)}} Q\left(\frac{T}{\sqrt{1+\sigma_i^2}}\right)$ and $\Gamma = 1 + T^2 + \sigma_i^2$ [21], [23].

It is clear from Fig. 4 that the output SNR of VOFDM always outperforms that of conventional OFDM even with a small number of VBs. It is also obvious that the simulation results of the latter system are in good agreement with the analytical ones, obtained from (17), which verifies the accuracy of our simulation model. Moreover, similar to the previous section, increasing the number of VBs will yield better output SNR performance.

In addition, it is seen that, for both nulling and clipping systems, when the threshold value is too small, the system performance deteriorates sharply as a result of the great loss in the useful signal energy. Similarly, when the threshold value is too high, performance will also degrade significantly. Hence, there exists an optimal threshold value that will maximize the output SNR of the systems under consideration. Notably, increasing the number of VBs will slightly reduce the optimal threshold. Furthermore, since the PLC channel is more accurately represented as frequency selective, we plot in Fig. 5 a sample of results for the output SNR versus the threshold value for both VOFDM and OFDM over the frequency selective PLC channel which is assumed to follow log-normal distribution [9], [29], [30], [33]. Comparing Fig. 4 and Fig. 5, it is clear that the channel frequency selective fading degrades the performance of both VOFDM and conventional OFDM systems similarly.

Next, the optimization problem of the threshold values is investigated more thoroughly.

## VI. PERFORMANCE OPTIMIZATION

Since our performance evaluation of the proposed VOFDM system is based on computer simulations, deriving mathematical expressions for the optimal tuning of the system parameters is beyond the scope of this paper. We therefore conduct in this section extensive computer simulations to find the optimal threshold values that will offer the maximum achievable output SNR and minimum bit error rate (BER) performances of the proposed system, for different values of the VBs and noise scenarios. This is obtained as

$$\begin{aligned}
\underset{T_i,\, i \in \{n,c\}}{\text{maximize}} \quad & \gamma_{o2}\left(T_i, M, p, \text{SINR}, \text{SNR}\right) \\
\text{subject to} \quad & 0 < T_i < 20\,\sigma_s^2 \\
& M = 1, 16, 32, 64
\end{aligned} \quad (20)$$

Clearly, equation (20) is a nonlinear objective function with a single-variable $(T_i)$. This optimization problem is solved numerically using the exhaustive search method. Now, using (20), we plot in Fig. 6 the maximum achievable output SNR of the proposed system, corresponding to the optimal nulling threshold values, with respect to SINR for $p = 0.01$ and $0.1$. In additional, the optimized output SNR curves of the conventional OFDM system are also presented on these plots.

For a fair comparison, the transmit power for both OFDM systems are assumed to be equal. With this in mind, it is evident from Fig. 6 that, at given $p$ and SINR values, the optimized VOFDM system can offer higher output SNR compared to conventional OFDM which means that lower transmit power values can be used in the proposed system and can still maintain same performance as the latter. In addition, comparing the results in Figs. 6a and 6b, it is clear that as the noise probability increases the achievable gains obtained with VOFDM become less significant. The other observation one can see is that for both noise probabilities, the highest gains are obtained in the intermediate SINR region. However, the two systems tend to have similar performance when SINR is very low.

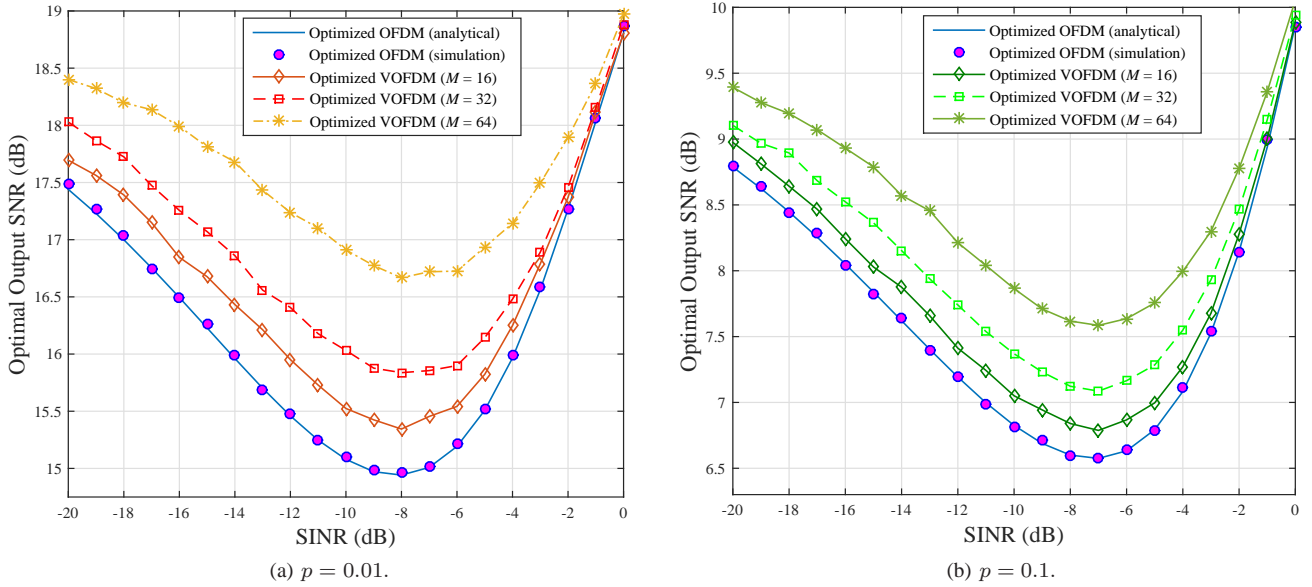

Figure 6: Maximum achievable output SNR performance as a function of SINR for the VOFDM-nulling system with various VB sizes and noise probabilities. Analytical and simulated output SNR results of the optimized conventional OFDM are also shown.

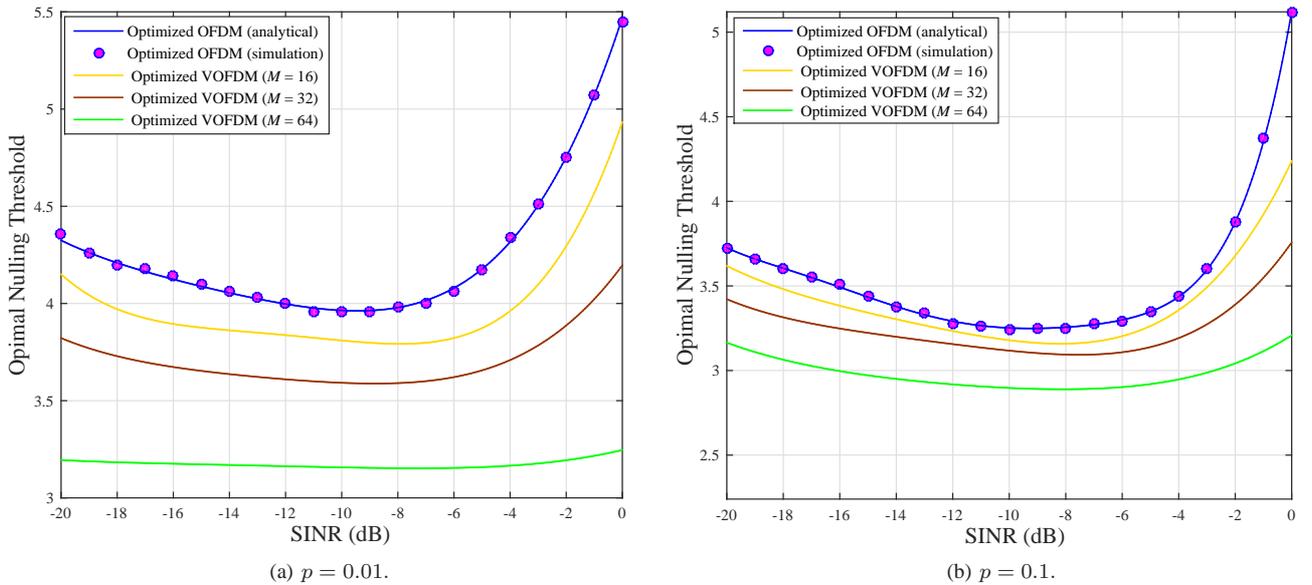

Figure 7: Optimal nulling threshold as a function of SINR for the VOFDM-nulling approach with various VB sizes and noise probabilities.

To better understand the behavior of these systems, we illustrate in Figs. 7a and 7b the optimal nulling thresholds corresponding to the output SNR curves shown in Figs 6a and 6b, respectively. It is interesting to notice that the optimal threshold for the proposed system decreases as the VB size is increased due to the fact that VOFDM lowers the PAPR, and hence lower nulling threshold will allow more efficient noise reduction. In addition, it is seen that conventional OFDM has the highest optimal threshold throughout the SINR spectrum. For all the systems considered here, the optimal thresholds are large when SINR is either extremely low or extremely high, which is more obvious in the conventional system. On the other hand, when SINR is very high, i.e. approaches 0 dB, the amplitudes of the noise pulses become very comparable to the information signal and therefore to avoid wrong nulling, large threshold values become optimal.

Now, comparing the results in Figs. 7a and 7b, we can see that the optimal threshold will be lower when the noise probability becomes more intensive. Also, interestingly enough, as the VB size is increased, the optimal threshold becomes less dependent on the noise characteristics and more so for small $p$ values. For instance, when $M = 64$, the optimal threshold becomes almost completely independent of SINR when $p = 0.01$, at around 3.2. This implies that if VOFDM is implemented with sufficiently large VB size with nulling at the receiver, it will be possible to always null the noise optimally by simply fixing the threshold value at 3.2. This can considerably reduce the complexity of the receiver compared to the case where estimating noise statistics is required to determine the optimal threshold. No doubt that this will further simplify the circuitry of the PLC modem, hence size and cost will be reduced.

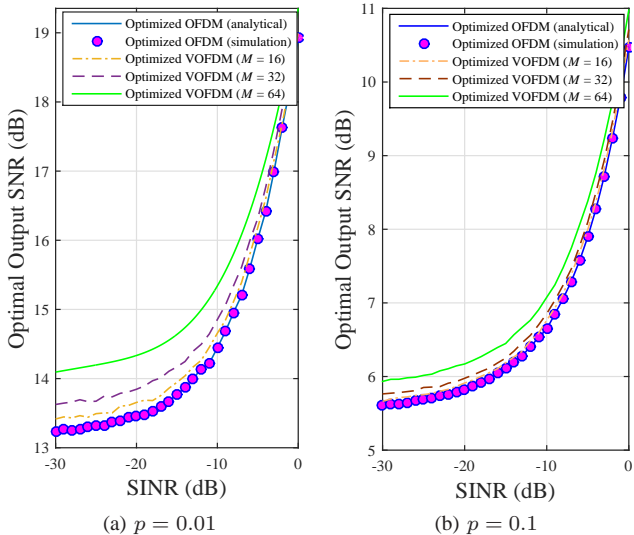

(a) $p = 0.01$

(b) $p = 0.1$

Figure 8: Maximum achievable output SNR performance versus SINR for the VOFDM-clipping approach with various VB sizes and noise probabilities. The output SNR results of the optimized conventional OFDM are also included.

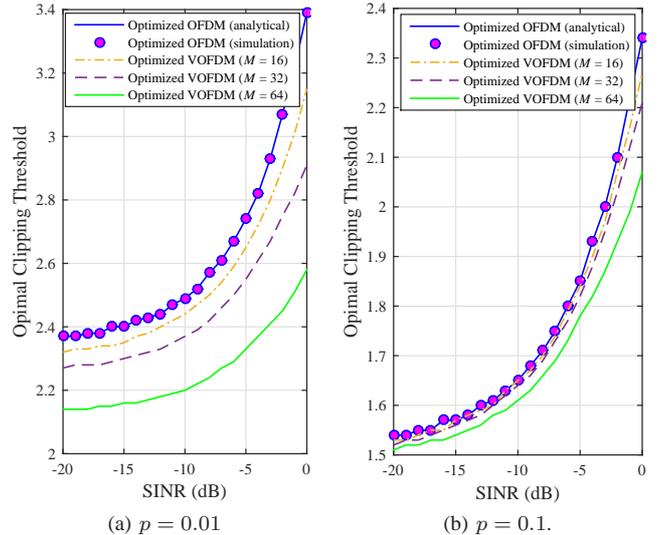

(a) $p = 0.01$

(b) $p = 0.1$.

Figure 9: Optimal clipping threshold value with respect to SINR for the VOFDM-clipping system with various VB sizes and noise probabilities. The output SNR results of the optimized conventional OFDM are also included.

On the other hand, Fig. 8 demonstrates the performance of both optimized VOFDM and conventional OFDM schemes when clipping is implemented, for two noise probabilities. Similar to the nulling case, VOFDM always outperforms conventional OFDM and this enhancement can be as high as 1 dB at low SINR values. Comparing Figs. 6 and 8, it can be noticed that unlike the nulling based systems in which performance enhances as SINR becomes smaller, in the clipping case, the performance worsens with reducing SINR. Further, the optimal clipping thresholds corresponding to the optimized output SNR curves in Fig. 8 are shown in Fig. 9. Unlike the optimal nulling threshold which levels off when $M$ is sufficiently large, the optimal clipping threshold always varies as the noise parameters are changed. Moreover, a sample of results of the minimum achievable BER performance corresponding to the optimal clipping threshold is shown in Fig. 10. Clearly, the proposed system always outperforms the conventional OFDM approach with clipping.

## VII. Practical Implementation of VOFDM

From the discussions above, VOFDM seems to be more suitable for the non-Gaussian PLC channel than conventional OFDM in terms of both cost and performance. The lower PAPR property of VOFDM will not only allow the deployment of cheaper nonlinear power amplifiers in PLC modems, but also, with some basic signal processing at the receiver such as nulling or clipping, can provide considerable transmit power savings. As a result, this transmit power reduction will reduce electromagnetic emissions from power lines. In terms of computational complexity, VOFDM seems to be more complex than conventional OFDM since it requires $M$ IFFTs and $M$ FFTs. However, this issue can be easily coped with in practice, thanks to the advances in processing chips available today with their low cost. In terms of hardware implementation complexity, the two OFDM systems do not differ significantly since they consist of the same main devices as demonstrated in Fig. 1. A brief comparison between the VOFDM and OFDM systems is illustrated in Table I. Finally, it is worth mentioning at this stage that practical implementation of the proposed system could not be done in this project due to the unavailability of the specialized hardware test-bed. This is now however a subject of future research.

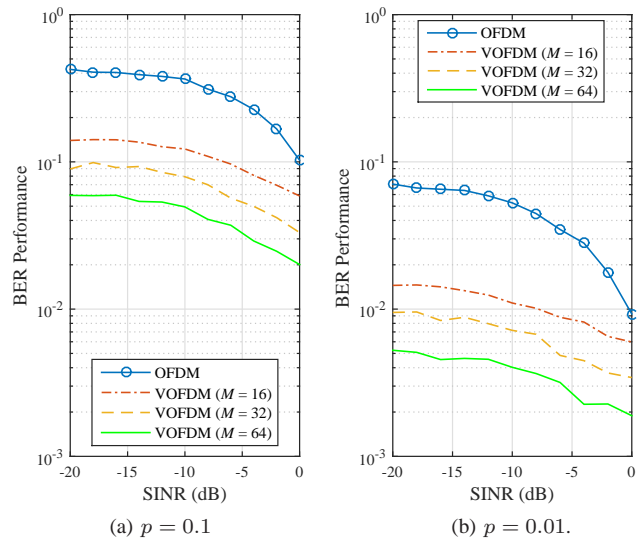

(a) $p = 0.1$

(b) $p = 0.01$.

Figure 10: Minimum achievable BER performance versus SINR for the VOFDM-clipping system with various VB sizes and noise probabilities.

## VIII. Conclusion

This paper proposed VOFDM for non-Gaussian PLC systems. The main advantage of VOFDM over conventional OFDM is its good PAPR property which becomes more important in non-Gaussian environments, such as the PLC network. It was shown that as we increase the number of VBs of the VOFDM system, the PAPR performance enhances and as a result noise detection becomes more accurate. Two nonlinear preprocessors, namely nulling and clipping, were implemented at the receiver and it was shown that optimizing the threshold value of the nonlinear devices is crucial to maximize performance.

| | OFDM | VOFDM |
|---|---|---|
| Suitability for PLC | Less suitable | More suitable |
| PAPR performance | Bad | Good |
| Transmitter complexity | Less complex | Complex |
| EMC | Bad | Good |
| Side information | No | No |

Table I: Comparison between VOFDM and conventional OFDM systems.

In general, VOFDM was found to be a promising scheme for PLC systems offering considerable transmit power savings compared to conventional OFDM. More specifically, under the same system setup, the VOFDM-nulling system can provide about 2 dB transmit power gains in comparison to conventional OFDM. This implies that power amplifiers with smaller dynamic range can be used, reducing cost and the EMC issue.

## Acknowledgment

This research has been carried out within the "Smart In-Building Micro Grid for Energy Management" project funded by EPSRC (EP/M506758/1) and supported by Innovate UK (Innovate UK Project 101836).